\documentclass[prl,twocolumn,aps,superscriptaddress,showpacs,showkeys,loatfix]{revtex4}
\usepackage{graphicx}
\usepackage{amsmath}

\begin{document}

\def\be{\begin{equation}}
\def\ba{\begin{eqnarray}}
\def\ee#1{\label{#1}\end{equation}}
\def\ea#1{\label{#1}\end{eqnarray}}
\def\la{\langle}
\def\ra{\rangle}
\def\bs{\begin{center}}
\def\es{\end{center}}
\def\fpa#1{\frac{\partial}{\partial #1}}
\def\ve{\varepsilon}
\def\De{D_{eff}}
\def\mv{\langle v \rangle}
\def\td{\textrm{d}}

\graphicspath{{img/}}

\title{Quantum diffusion in biased washboard potentials: strong
  friction limit}

\author{L. Machura}
\affiliation{Institute of Physics, University of Augsburg,
D-86135 Augsburg, Germany}
\affiliation{Institute of Physics, University of Silesia,
40-007 Katowice, Poland}
\author{M. Kostur}
\affiliation{Institute of Physics, University of Augsburg,
D-86135 Augsburg, Germany}
\author{P. Talkner}
\affiliation{Institute of Physics, University of Augsburg,
D-86135 Augsburg, Germany}
\author{J. \L uczka}
\affiliation{Institute of Physics,  University of Silesia,
40-007 Katowice, Poland}
\author{P. H\"anggi}
\affiliation{Institute of Physics, University of Augsburg,
D-86135 Augsburg, Germany}
\date{\today}
\begin{abstract}
  Diffusive transport properties of a quantum Brownian particle moving
  in a tilted spatially periodic potential and strongly interacting
  with a thermostat are explored. Apart from the average stationary
  velocity, we foremost investigate the diffusive behavior by
  evaluating the effective diffusion coefficient together with the
  corresponding Peclet number. Corrections due to quantum effects,
  such as quantum tunneling and quantum fluctuations, are shown to
  substantially enhance the effectiveness of diffusive transport if
  only the thermostat temperature resides within an appropriate
  interval of intermediate values.
\end{abstract}
\maketitle


\section{Introduction}

Brownian motion in periodic structures can describe diverse processes
in many different branches of science. Within a physical context,
among other phenomena, it models the dynamics of the phase difference
across a Josephson junction \cite{junction,schonzaikin}, rotating
dipoles in external fields \cite{Reg2000, Coffey}, superionic
conductors \cite{Ful1975}, charge density waves \cite{Gru1981},
particle separation by electrophoresis \cite{Ajd1992}, transport on
crystalline surfaces \cite{kat2005,marchesoni05}, biophysical
processes such as intracellular transport \cite{BM,
  RMP,Bie2003,Lin2002}. Yet another important area constitutes the
noise-assisted transport of Brownian particles \cite{RevMod,Melnikov},
as it occurs for Brownian motors possessing ample applications in
physics and chemistry \cite{BM}.

In this paper, we study the one-dimensional overdamped motion of a
quantum Brownian particle subjected to a tilted potential $U(x)$,
\begin{equation}\label{U}
U(x) = V(x) - Fx, \quad V(x) = V(x+L),
\end{equation}
where $V(x)$ denotes a periodic potential of period $L$ and $F$ is
an external static force.

The basic quantities characterizing this motion are statistical
moments of position and velocity of the Brownian particle.  At least
the first two moments i.e. the average position and average velocity
and their respective dispersions are most substantial.  In particular,
the stationary average velocity can be defined by the relation
\be
\mv  = \lim_{t \to \infty} \frac{\la x(t) \ra }{ t},
\ee{meanv}
where $x(t)$ is the position of the Brownian particle at time $t$ and
$\la \ldots \ra$ is the average over all realizations of the thermal
noise and the initial conditions.  The dispersion of the position can
be characterized by the diffusion coefficient defined as
\cite{JPCMach}
\be
\De = \lim_{t \to \infty} \frac{\la x^2(t) \ra - \la x(t) \ra^2}{2t}.
\ee{deff}
In the classical case, for strong friction, when only thermal
equilibrium fluctuations affect the particle, the stationary average
velocity and the diffusion coefficient can be expressed by exact
closed formulas \cite{reimann1,reimann2}.  For nonequilibrium driving
they can be calculated in specific cases only, see e.g.
\cite{JLepl,dubkov,spagnolo}.

Depending on the form of the potential, the magnitude of the tilt and
thermostat temperature, two interesting phenomena can be observed: A
giant enhancement of the diffusion constant at a critical tilt
\cite{reimann1,reimann2} and a low randomness window at sub-critical
tilts \cite{lindner}. The influence of the shape of the potential
\cite{Hei2004c,lindner} and of a position dependent friction
coefficient \cite{dan} on the transport have also been investigated.

\begin{figure}[htbp]
\bs
\includegraphics[scale=.4]{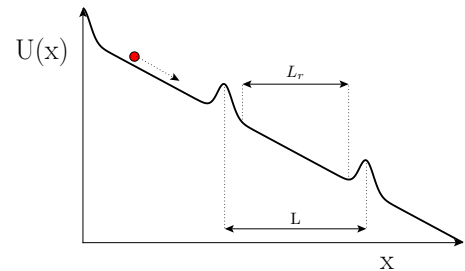}
\caption{Schematic picture of the Brownian particle in the tilted
  periodic potential $U(x)=V(x)-Fx$, defined in (\ref{U}) and
  (\ref{v}) below, plotted for the following set of parameters:
  $\Delta = 10$, $\ve = 10$, $F=0.2$, $L$ denotes the period the
  potential $V(x)$ and $L_r$ stands for the length of free slide
  between the barriers. }
\label{fig1}
\es
\end{figure}

However, in many cases, like for the Josephson junction at
intermediate temperatures, the classical theory is insufficient; i.e.,
the leading quantum corrections \cite{ingold85} should be considered.
It was shown in Ref. \cite{ankerhold, anker2} that in the strong
friction limit, effects of quantum Brownian fluctuations
\cite{hanggi05} are restricted not only to low temperatures;
therefore, these should be incorporated for higher temperatures as
well. This is so because quantum fluctuations, even if reduced for one
variable, are enlarged for the conjugate variable. Quantum corrections
can modify the dynamics quantitatively and sometimes even
qualitatively. Physically relevant examples illustrating these
features are presented in Refs. \cite{anker2,hanggi05,machura}.

For the average current in (\ref{meanv}) of the Brownian particle
dynamics such quantum corrections have been studied repeatedly in the
previous literature within different approaches \cite{QCV}. In
particular, within the quantum Smoluchowski equation these quantum
corrections have been studied recently by Ankerhold in Ref.
\cite{ankerholdJ} for an overdamped Josephson junction.

In distinct contrast, with this work we mainly focus on the role of
the quantum corrections to the diffusion of the mean squared
displacement of the coordinate (or Josephson- phase, respectively) as
defined with (\ref{deff}).

The quantum diffusive dynamics in the regime of strong interaction
with a thermostat can be described by the so called Quantum
Smoluchowski Equation (QSE) \cite{ankerhold, anker2} incorporating
quantum fluctuations above the crossover temperature \cite{ingold85,
  RevMod}. It corresponds to a classical Smoluchowski equation, in
which the potential $U(x)$ and the diffusion coefficient $D$ are
modified due to quantum effects. In other words, the quantum
non-Markovian diffusion process of a particle position is approximated
by a classical Markov process describing a motion in an effective
potential and with an effective, state-dependent diffusion
coefficient. This leads to a quite comfortable situation because
methods of analyzing Smoluchowski equations are well elaborated and
can directly be applied and implemented for a wide class of systems.
In this paper we demonstrate how the quantum fluctuations influence
the diffusion behavior of a Brownian particle stochastically moving in
washboard-like potentials.


\section{Classical Brownian particle in  tilted periodic potentials}

The overdamped motion of a classical Brownian particle is described by
the Langevin equation
\begin{equation}\label{LaCl}
\eta \dot{x} = -U'(x)  + \sqrt{2 \eta kT}~\xi(t),
\end{equation}
where $\eta$ denotes the viscous friction coefficient and $k$ is the
Boltzmann constant. The dot and the prime indicate differentiation
with respect to time $t$ and to position $x$, respectively. The
zero-mean and $\delta$-correlated Gaussian white noise $\xi(t)$ models
the influence of a thermostat of temperature $T$ on the system.

The calculation of the average velocity $\mv$ and effective diffusion
$\De$ can be accomplished by mapping the washboard potential on a
corresponding jump process. This construction procedure has been
elucidated in \cite{lindner}. As a result, a cumulative process with
independent increments is obtained and its asymptotic mean and
variance are given by the first two central moments of the escape time
density \cite{RevMod, reimann1,reimann2}.  In this way, the stationary
average velocity \cite{RevMod,Melnikov, strat, tilted} and the
diffusion coefficient \cite{reimann1,reimann2} for the process
modelled by (\ref{LaCl}) can be expressed by closed form relations
involving quadratures only; i.e.,
\ba
 \mv & = & \frac{L}{T_1(x_0\to x_0+L)},\\ \nonumber
\De & = & \frac{L^2}{2} \: \frac{T_2(x_0\to x_0+L) - T_1^2(x_0\to
x_0+L)} {T_1^3(x_0\to x_0+L)}, \ea{clvD} where $x_0$ is an
arbitrary, initial value  and
\ba
T_n(x_0\to b) = \la t^n(x_0\to b)\ra
\ea{Tn}
denotes the $n$th statistical moment of the first passage time
$t(x_0\to b)$ at which the Brownian particle arrives at the point $b$
while starting out from the position $x_0$. For the case $b> x_0$,
these moments are given by the recurrence relation \cite{Mironov},
\ba T_n(x_0\to b) = n \beta \eta \int_{x_0}^b \td x \: \exp[\beta
U(x)]
\nonumber\\
\times \int_{-\infty}^x \td y \: \exp[-\beta U(y)] T_{n-1}(y\to b)
\ea{mom} for $n=1,2,3...$, where $T_{0}(y\to b)=1$,  $\beta =1/kT$
and the product $\beta \eta = D_0^{-1}$ is the reciprocal of the
Einstein diffusion coefficient $D_0$. The expressions (\ref{clvD})
and (\ref{mom}) are rather complicated. However, they can be
simplified  as shown in \cite{reimann1,reimann2}.


\section{Overdamped quantum Brownian motion}

To start with the investigation of quantum corrections to diffusion we
consider a quantum Brownian particle moving in the tilted potential
$U(x)$. The evolution of its position can be described by the
respective probability density function $P(x,t)=\la x \vert \rho (t)
\vert x \ra$, which is the diagonal part of the statistical operator
$\rho (t)$.  Within the strong friction limit (the quantum
Smoluchowski regime), the dynamics of such a particle is described by
the Quantum Smoluchowski Equation (QSE) that takes into account
leading quantum corrections.  It has the structure of a classical
Smoluchowski equation with modified drift and modified diffusion terms
\cite{ankerhold,anker2,machura}
\begin{equation}\label{FP}
\eta \frac{\partial}{\partial t} P(t,x)= \frac{\partial}{\partial
x}\Big(U_{eff}'(x) + \frac{\partial }{\partial x} D(x)\Big)P(x,t)\;.
\end{equation}
The effective potential reads \be U_{eff}(x) = U(x) + (1/2) \lambda
U''(x). \ee{veff}
The effective diffusion coefficient $D(x)$, being constant in the
classical case, i.e., $D(x)= D = k_B T = \beta^{-1}$, becomes
position-dependent, assuming the unique form \cite{machura,rudnicki},
\be D(x) = \Big( \beta [ 1 - \lambda \beta U''(x) ] \Big)^{-1}.
\ee{qdeff} 
This diffusion is required to remain non-negative, i.e., within its
regime of validity \cite{machura,rudnicki}, the inequality $\lambda
\beta U''(x) = \lambda \beta V''(x) < 1$ must be satisfied for all
positions $x$. For smooth periodic functions $V(x)$ and sufficiently
small $\lambda\beta$ this inequality holds for arbitrary $x$.

The prominent parameter $\lambda$ characterizes quantum fluctuations
in position space; it explicitly reads \cite{ankerhold,anker2},
\be \lambda = (\hbar/ \pi \eta) \ln (\hbar \beta \eta / 2 \pi M)\;.
\ee{lambda}
It depends nonlinearly on the Planck constant $\hbar$ and on the mass
$M$ of the Brownian particle, whereas, in the classical case, the
overdamped dynamics does neither depend on $\hbar$ nor on the mass $M$
(note that we use the friction constant $\eta$ which has the unit
$[kg/s]$ as in the classical Stokes case). Note also that this quantum
correction approaches zero with the friction $\eta$ growing towards
infinity.

The Langevin equation corresponding to the Smoluchowski equation
(\ref{FP}) becomes within the Ito-interpretation \cite{PR},
\begin{equation}\label{LE}
\eta \dot{x} = - U_{eff}'(x)  + \sqrt{2\eta  D(x)}~\xi(t) \;.
\end{equation}

The average stationary velocity $\mv$ and the diffusion classical
coefficient $\De$ can be calculated as in the case described by the
Langevin equation (\ref{LaCl}), using the relations (\ref{clvD}) and
the known formula for statistical moments of the first passage time.
In comparison with (\ref{mom}), the statistical moments are thereby
modified into the form \cite{Mironov}
\ba T_n(x_0\to b) = n \eta \int_{x_0}^b \td x \: \exp[\phi (x)]
\nonumber\\
\times \int_{-\infty}^x  \td y \:  D^{-1}(y) \exp[-\phi (y)]
 T_{n-1}(y \to b)\;,
\ea{mom2}
where
\ba
\phi (x)  =  \int^x  \frac{V'_{eff}(z)}{ D(z)} \: \td z.
\ea{phi}
Insertion of  the expressions for the effective potential
(\ref{veff}) and the effective diffusion function (\ref{qdeff})
yields
\ba T_n(x_0\to b) = n \beta \eta \  \int_{x_0}^b \td x \:
\exp[\beta\psi(x)]
\nonumber\\
\times \int_{-\infty}^x  \td y \:  \exp[-\beta \psi(y)]
\left[1-\lambda \beta U''(y)\right]
 T_{n-1}(y \to b),
\ea{mom3}
where the thermodynamic potential $\psi(x)$ becomes
\ba
\psi (x) & = & U(x) +(1/2) \lambda U''(x) +\\\nonumber
& - & (1/2) \lambda \beta [U'(x)]^2 -(1/4)\lambda^2 \beta [U''(x)]^2.
\ea{psi}
We observe that quantum corrections modify the statistical moments as
given by eq. (\ref{mom3}) compared to the classical form (\ref{mom})
in a two-fold way: First, the physical potential $U(x)$ is replaced by
the thermodynamic potential $\psi(x)$. This thermodynamic potential
depends on the temperature $\beta$ of the system and on the coupling
constant of the Brownian particle with its surroundings via the
damping constant $\eta$, which in turn enters into the parameter
$\lambda$.  Second, the function in the inner integral (over the
variable $y$) on the right hand side of eq. (\ref{mom3}) is modified
by the factor $ [1-\lambda \beta U''(y)]$, which depends on the
curvature (i.e. on $U''(y)= V''(y)$) of the physical potential $U(y)$.

\begin{figure}[htbp]
\bs
\includegraphics[scale=.7]{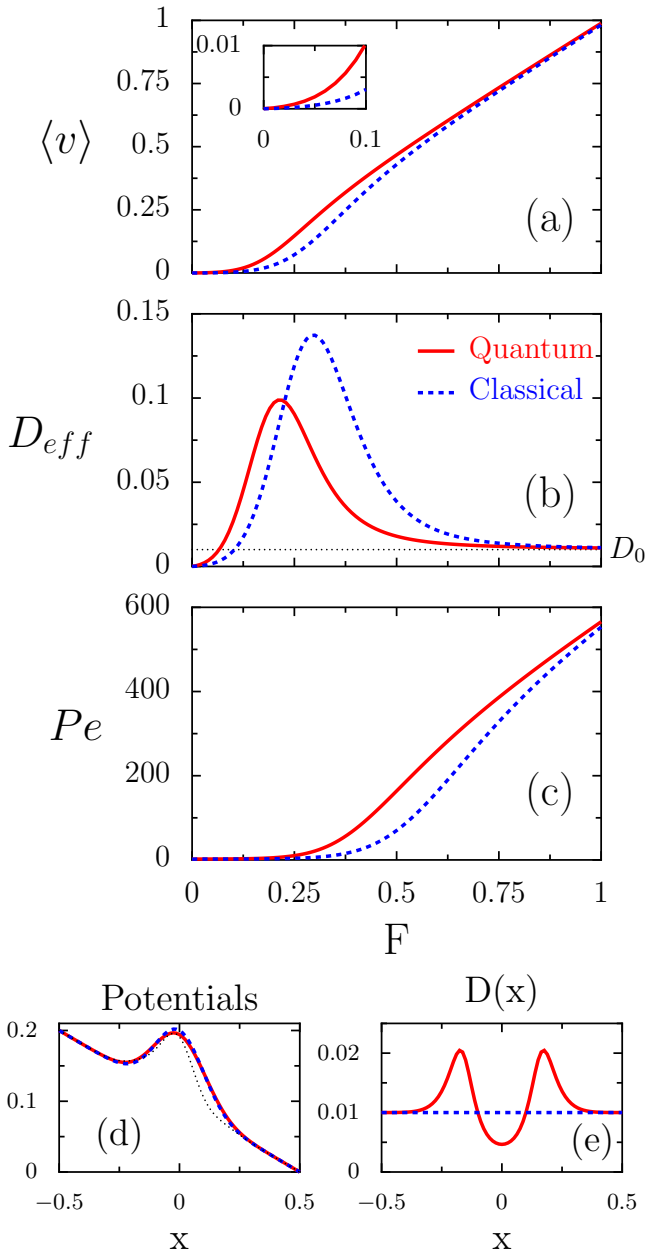}
\caption{ Average velocity $\la v \ra$ (a), effective diffusion
  $D_{eff}$ (b) and Peclet number $Pe$ (c) are drawn versus tilt $F$.
  Parameters of the potential are: $\ve=100$, $\Delta = 10$, which
  result in a barrier height of $\Delta V = 0.1$ in the absence of a
  tilt ($F=0$), and a critical tilt $F_{c}\simeq0.6$ above which all
  barriers disappear. The rescaled inverse temperature is set to
  $\beta = 100$, corresponding to a small quantum parameter $\lambda_0
  \simeq 0.00115129$. The mean velocity and Peclet number are always
  larger in a quantum case, even if the classical effective diffusion
  is smaller (as is the case for $F \lesssim 0.25$). With the panel
  (d) we depict the corresponding classical, $U(x)$, (dashed), quantum
  effective, $U_{eff}(x)$, (solid line) and thermodynamic, $\psi(x)$,
  (dotted) tilted potentials, respectively. The position-dependent
  diffusion coefficient is plotted in panel (e). } \label{fig2} \es
\end{figure}

We will examine the quantum Brownian particle moving in a tilted
washboard potential like that presented in Fig. \ref{fig1}. We scale
the force and the diffusion coefficient in such a way that for the
rescaled equation corresponding to (\ref{LE}) the friction coefficient
equals $\eta =1$. We choose a specific form of the periodic part
$V(x)$ of the potential $U(x)$, namely \cite{lindner}:
\be V(x) = \Delta \: \exp[\varepsilon (\cos(x) - 1)]/ \varepsilon
\;. \ee{v}
The advantage of this choice is that by an appropriate manipulation of
$\ve$ and $\Delta$, the barrier height and the distance between
neighboring barriers can be varied independently. As a consequence,
one can change two time-scales independently: a first one is related
to the deterministic sliding motion between neighboring barriers and
the other is related to the inverse of the activation rate over a
barrier.


\section{Quantum diffusion  in tilted  washboard potentials}
In order to understand the influence of the shape of the washboard
potential on the particle dynamics, both in the classical and quantum
regimes, is it desirable to identify characteristic time-scales.  The
first one is given by the time $\tau_r = L_r/F$ the particle needs to
slide down the distance $L_r$ (see in Fig.\ref{fig1}) with a constant
velocity $v=F$ (remember that the friction coefficient $\eta =1$ and
the force is rescaled).  This timescale is relevant if the potential
has an almost constant slope between neighboring comparatively narrow
barriers, as in the case considered here.  The second time-scale
$\tau_e$ is determined by the escape time over the barrier.  The
potential $V(x)$ has been chosen in the above described way, so that
these time-scales can be 'tuned' independently: $\tau_r$ by the force
$F$ and the parameter $\ve$, and $\tau_e$ by the barrier height using
both parameters $\ve$ and $\Delta$ of the potential (\ref{v}).
\begin{figure}[htbp]
\bs
\includegraphics[scale=.6]{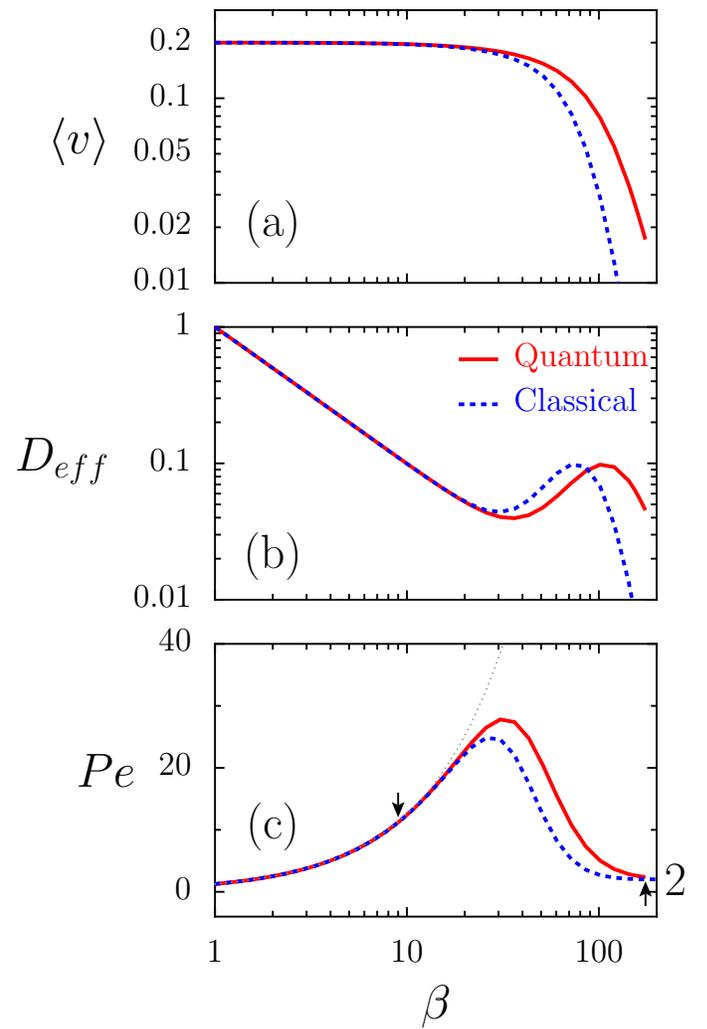}
\caption{ Average velocity (a), effective diffusion (b) and Peclet
  number (c) versus the inverse temperature $\beta$.  The chosen
  parameters are: bias $F = 0.2$, potential parameters
  $\varepsilon=100$ and $\Delta = 10$.  The two arrows in panel (c)
  mark the characteristic temperatures where the system starts to
  markedly deviate from the classical behavior (see text for details).
  The region of intermediate temperatures is located between the
  downward and upward pointing arrows.  It bridges between the regions
  of freely sliding Brownian particle dynamics, characterized by the
  Peclet number $Pe = F L \beta$ (see the thin, dotted line in (c))
  and the Poisson like behavior with $Pe=2$.  Within this region the
  Peclet number assumes its maximal value.}
\label{fig3}
\es
\end{figure}

The transport of particles is optimal if a large mean velocity goes
along with small diffusion. This can be quantified by the
dimensionless Peclet number \cite{peclet}:
\be Pe = \frac{\mv L}{\De}.
 \ee{pe}
 The efficiency of the diffusive transport as measured by the Peclet
 number can either be enhanced by an increase of the net current (i.e.
 the stationary mean velocity) and/or by a decrease of the effective
 diffusion, resulting in a maximal Peclet number $Pe$. The average
 velocity for the overdamped motion in a tilted washboard potential is
 limited by the free-slide speed which coincides with the value $F$ in
 our case. The particle will approach this free-slide velocity when
 the barriers become negligible, for example for a sufficiently high
 thermal energy $kT$ or very strong tilt $F$. This situation, however,
 does not lead to an optimal transport performance in the sense of a
 maximal Peclet number \cite{lindner,MachuraJCP}, see also below.

\begin{figure}[htbp]
\bs
\includegraphics[scale=1]{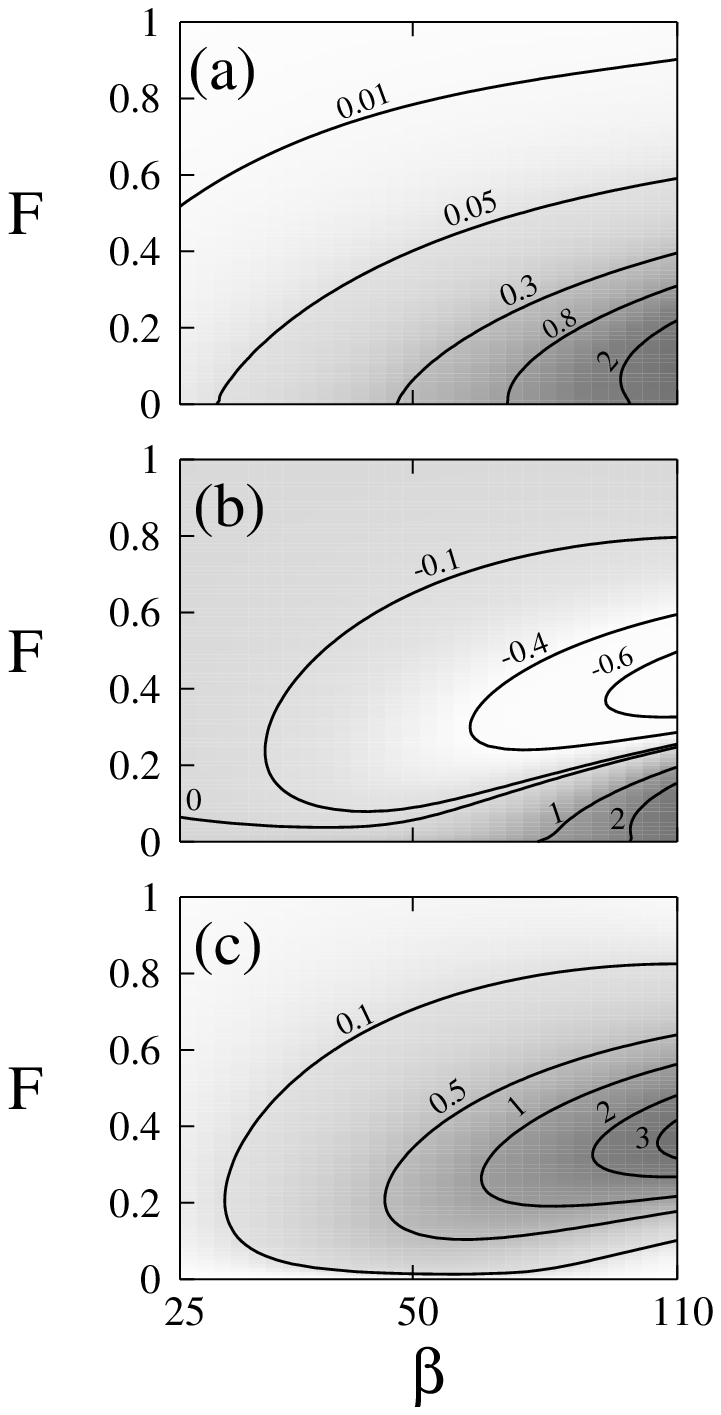}
\caption{The influence of quantum corrections are illustrated as a
  function of bias $F$ {\it vs.} inverse temperature $\beta$ by the
  relative difference between the quantum ($Q$) and the corresponding
  classical ($C$) values of the current $(\mv^Q - \mv^C)/\mv^C$ (a),
  the effective diffusion $(\De^Q - \De^C)/\De^C$ (b) and the Peclet
  number $(Pe^Q - Pe^C)/Pe^C$ (c), respectively. The positive values of
  the depicted quantities indicate the relevant role of quantum effects.}
\label{fig4}
\es
\end{figure}

\subsection{Quantum-renormalization of the barrier shape}
We studied the influence of quantum corrections on transport in tilted
periodic systems by means of a numerical analysis of the basic
expressions (\ref{clvD}) and (\ref{mom3}). We found that the quantum
current is always higher than the corresponding classical one (see
Figs. \ref{fig2}, \ref{fig3} and \ref{fig4}).  This phenomenon can be
explained by comparing the potential $U(x)$, the effective quantum
potential $U_{eff}(x)$ and the thermodynamic potential $\psi(x)$ with
each other, as well as by analyzing the effective diffusion function
$D(x)$ (which is constant in the classical case), see Fig. \ref{fig2}.
Clearly, the effective quantum potential $U_{eff}(x)$ (solid line)
possesses slightly lower and thinner barriers than $U(x)$ (dashed
line).

The state-dependent diffusion function $D(x)$ possesses maxima and
minima.  The maxima, which are shifted away from the potential barrier
locations, can be interpreted as a higher effective local temperature.
The minimum of $D(x)$ is located in the neighborhood of the top of the
barrier (near $x\sim 0$ in panel (e)). It means that quantum
fluctuations mimic an effective temperature which is lower at the
barrier and higher in the potential wells. For the escape dynamics the
thermodynamic potential $\psi(x)$ is decisive: It contains the
combined influences of the effective potential and the effective
diffusion. In the present case, $\psi(x)$ displays both a lower and a
narrower barrier than the bare potential $U(x)$ of the corresponding
classical process.  It is remarkable that for all cases considered,
the Peclet number is always larger in the quantum case than in the
classical case, thus providing a more coherent motion. This behavior
is exemplified in Figs. \ref{fig2}, \ref{fig3} and \ref{fig4}.

\subsection{Role of  quantum corrections for diffusion}
Depending on the relation between the thermal energy $kT$ and the
barrier height of the tilted potential, one can distinguish regimes of
high and low temperatures, denoted in Fig. \ref{fig3} by a downward
and an upward arrow, respectively. To the left of downward arrow, i.e.
for high temperatures, the particle barely feels the barriers and thus
freely slides down the hill with an average velocity given by $\mv
\approx F$.

The effective diffusion coefficient $\De$ can then be approximated by
the Einstein diffusion coefficient $1/\beta$ in both the classical and
in the quantum cases.  The corresponding approximated values of the
Peclet number, given by $Pe = F L \beta$, are depicted as a dotted
line.  To the right of the upward arrow in Fig.  \ref{fig3}c the
relaxation time $\tau_r$ is much smaller than the time scale for
barrier crossing.  The time evolution in that case consists of a
sequence of independent activations. Indeed, the value of Peclet
number approaches $Pe = 2$ which is characteristic for Poisson
process.
\begin{figure}[htbp]
\bs
\includegraphics[scale=0.5]{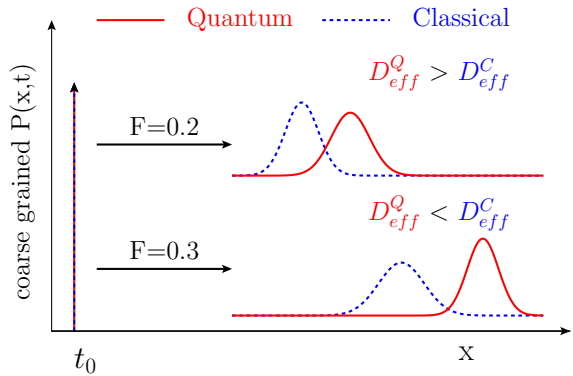}
\caption{ The influence of the quantum corrections on the evolution of
  the coarse grained probability density function $P(x,t)$ (see
  discussion in \cite{lindner}) is illustrated for the same set of
  parameters as in Fig. \ref{fig2} and for different forces $F=0.2$
  and $0.3$.  Note that in the quantum case the velocity is always
  {\it larger}; therefore, the maximum of the density is located at a
  more distant position as compared to the classical case. The width
  of the density differs depending on the driving parameters and
  thereby reflects the effective diffusion strength.}
\label{fig5}
\es
\end{figure}

The most interesting region is located between the two temperatures,
indicated by two oppositely directed arrows (one downwards and one
upwards), where we found the optimal transport, i.e. the maximum of
the Peclet number $Pe$.  The quantum behavior significantly deviates
from the classical one only within this very region. We observe that
the transport quality, expressed in terms of the characterizer $Pe$ is
never suppressed by quantum effects (see Fig. \ref{fig3} and
\ref{fig4}), even though quantum corrections may increase the the
effective diffusion $D_{eff}$.

In Fig. \ref{fig4} we depict the relative value for the correction of
$D_{eff}$ and $Pe$, respectively, in the parameter space spanned by
$\beta,F$.  First, it is detectable that the corrected values of
velocity, the effective diffusion and Peclet number may differ up to
$200\%$ from the classical values. Second, the sign of the relative
corrections of $\mv$ and $Pe$ is positive, but in the case of the
effective diffusion $\De$ it might assume negative values within some
range of parameters.

Finally, we illustrate the effect of quantum corrections on the
transport in tilted washboard potential. In Fig. \ref{fig5} we present
two examples of the time evolution of the density function $P(x,t)$.
The impact of quantum corrections is clearly visible. The width of
$P(x,t)$ becomes larger at $F=0.2$ and smaller at $F=0.3$ when the
quantum corrections are acting, but the peak of the probability
density travels in the quantum case with a larger velocity in both
situations.

\section{Conclusions}
The effect of the quantum contribution on Brownian motion, in
particular on the diffusion of particles and the related transport
performance is addressed in this work.  The quantum current always
exceeds the corresponding classical one; -- quantum features like
tunneling and quantum fluctuations seemingly always assist the
particle to overcome barriers and to pass longer distances, resulting
in larger average stationary velocity.  The diffusion coefficient,
$\De$, is found to assume, generally, a non-monotonic function of the
static force $F$ and the temperature $\beta$.  In other words --
optimal conditions exist for both, directed and diffusive transport.
Depending on the parameters of the system, quantum effects may either
increase or decrease the effective diffusion of the particle. The
Peclet number is found to be {\it always} larger for quantum systems,
-- see in Fig.  \ref{fig4} (c).  The most significant finding is that
quantum effects always improve the diffusive quantum transport for the
class of systems considered in this work.

\section*{Acknowledgment}
The authors gratefully acknowledge helpful discussions with Gerhard
Schmid. The work was supported by the Deutsche Forschungsgemeinschaft
via grant HA 1517/13-4, the Graduiertenkolleg 283 (LM), the
collaborative research grant SFB 486 and the DAAD-KBN (German-Polish
project {\it Stochastic Complexity}) (PH and J{\L}).


\begin{thebibliography}{99}

\bibitem{junction} A. Barone and G. Patern\`o, {\it Physics and
    Application of the Josephson Effect}, (Wiley, New York, 1982).

\bibitem{schonzaikin} G. Sch\"on and A. D. Zaikin, Phys. Rep. {\bf
    198}, 2378 (1990).

\bibitem{Reg2000} D. Reguera, J. M. Rubi, and A. P\'erez-Madrid, Phys.
  Rev. E {\bf 62}, 5313 (2000); D. Reguera, P. Reimann, P. H\"anggi
  and J. M. Rubi, Europhys. Lett. {\bf 57}, 644 (2002).

\bibitem{Coffey} W. T. Coffey, Yu. P. Kalmykov and J. T. Waldron, {\it
    The Langevin Equation}, 2-nd edition, (World Scientific,
  Singapore, 2004) see sects. 5 and 7-10 therein.

\bibitem{Ful1975} P. Fulde, L. Pietronero, W.~R. Schneider, and S.
  Str\"assler, Phys.  Rev. Lett. {\bf 35}, 1776 (1975); W. Dieterich,
  I. Peschel, and W.~R. Schneider, Z. Physik B {\bf 27}, 177 (1977);
  T. Geisel, Sol.  State Commun. {\bf 32}, 739 (1979).

\bibitem{Gru1981} G. Gr\"uner, A. Zawadowski, and P. M. Chaikin, Phys.
  Rev. Lett. {\bf 46}, 511 (1981).

\bibitem{Ajd1992} A. Ajdari and J. Prost, Proc. Natl. Acad. Sci. USA.
  {\bf 88}, 4468 (1991).

\bibitem{kat2005} E. Pollak, J. Bader, B. J. Berne and P. Talkner,
  Phys. Rev. Lett.  {\bf 70}, 3299 (1993); M. Borromero and F.
  Marchesoni, Surf. Sci.  {\bf 465}, L771 (2000).

\bibitem{marchesoni05} M. Borromeo and F. Marchesoni, Chaos {\bf 15},
  026110 (2005).

\bibitem{BM} P. H\"anggi and R. Bartussek, Lect. Notes Phys.
  \textbf{476}, 294 (1996); R. D. Astumian and P. H\"anggi, Physics
  Today \textbf{55} (11), 33 (2002); P. Reimann and P. H\"anggi, Appl.
  Phys. A \textbf{75}, 169 (2002); P. Reimann, Phys. Rep.
  \textbf{361}, 57 (2002); H. Linke, Appl. Phys. A \textbf{75}, 167
  (2002); P. H\"anggi, F. Marchesoni, F. and Nori, Ann. Phys.
  (Leipzig) {\bf 14}, 51 (2005).

\bibitem{RMP} F. J\"ulicher, A. Ajdari, J. Prost, Rev. Mod. Phys.
  \textbf{69}, 1269 (1997); E.~Frey, K. Kroy, Ann. Phys. (Leipzig)
  \textbf{14}, 20 (2005); J.~Howard, {\it Mechanics of Motor Proteins
    and the Cytoskeleton}, (Sinauer Assoc., Sunderland, 2001).

\bibitem{Lin2002} B. Lindner, L. Schimansky-Geier, Phys. Rev. Lett.
  {\bf 89}, 230602 (2002).

\bibitem{Bie2003} M. Bier, Phys. Rev. Lett. {\bf 91}, 148104 (2003).

\bibitem{RevMod} P. H\"anggi, P. Talkner, M. Borkovec, Rev. Mod. Phys.
  {\bf 62}, 251 (1990).

\bibitem{Melnikov} V. I. Melnikov, Phys. Rep. {\bf 209}, 1 (1991).

\bibitem{JPCMach} A discussion about the equivalence of the different
  definition of $D_{eff}$ can be found in the Appendix in: L. Machura
  {\it et al.}, J. Phys.: Condens. Matter {\bf 17}, S3741 (2005).

\bibitem{reimann1} P. Reimann, C. Van den Broeck, H. Linke, P. H\"anggi,
J. M. Rubi and A. P\'erez-Madrid, Phys. Rev. Lett. {\bf 87}, 010602
(2001).

\bibitem{reimann2} P. Reimann, C. Van den Broeck, H. Linke, P. H\"anggi,
J. M. Rubi and A. P\'erez-Madrid, Phys. Rev. E {\bf 65}, 031104
(2002).

\bibitem{JLepl} J. \L uczka, R. Bartussek and P. H\"anggi, Europhys.
  Lett. {\bf 31}, 431 (1995); P. H\"anggi, R. Bartussek, P. Talkner,
  J. \L uczka, Europhys. Lett. {\bf 35}, 315 (1996); J. Kula, T.
  Czernik and J. \L uczka, Phys. Lett. A {\bf 214}, 14 (1996).
\bibitem{dubkov} A. N. Malakhov, Pisma w JETP 24, 9 (1998) [Techn.
  Phys. Lett {\bf 24}, 833 (1998)]; A. A. Dubkov, Pisma w JETP 29, 18
  (2003) [Techn. Phys. Lett. {\bf 29}, 92 (2003)].

\bibitem{spagnolo}
B. Spagnolo, A. A. Dubkov, and N. V. Agudov, Physica A {\bf 340},
265 (2004).

\bibitem{lindner} B. Lindner, M. Kostur and L. Schimansky-Geier,
  Fluct. Noise Lett. {\bf 1}, R25 (2001).

\bibitem{Hei2004c} E. Heinsalu, T. Ord, and R. Tammelo, Phys. Rev. E
  {\bf 70}, 041104 (2004); T. Ord, E. Heinsalu, and R. Tammelo, Eur.
  Phys. J. B {\bf 47}, 275 (2005).

\bibitem{dan} D. Dan and A. M. Jayannavar, Phys. Rev. E {\bf 66},
  041106 (2002).

\bibitem{ingold85}
P. H\"anggi, H. Grabert, G. L. Ingold and U. Weiss, Phys. Rev. Lett.
{\bf 55}, 761 (1985).

\bibitem{ankerhold} J. Ankerhold, P. Pechukas, and H. Grabert, Phys.
  Rev. Lett. {\bf 87}, 086802 (2001).

\bibitem{anker2} J. Ankerhold, H. Grabert and P. Pechukas, Chaos {\bf
    15}, 026106 (2005).

\bibitem{hanggi05} P. H\"anggi and G. L. Ingold, Chaos {\bf 15},
  026105 (2005).

\bibitem{machura} L. Machura, M. Kostur, P. H\"anggi, P. Talkner and
  J. {\L}uczka, Phys. Rev. E {\bf 70}, 031107 (2004).

\bibitem{QCV} V. I. Melnikov and A. S\"ut\"o, Phys. Rev. B {\bf 34},
  1514 (1986); W. Zwerger, Phys. Rev. B {\bf 35}, 4737 (1987); P.
  Reimann, M.  Grifoni and P. H\"anggi, Phys. Rev. Lett. {\bf 79}, 10
  (1997); H.  Grabert, G. L. Ingold and B. Paul, Europhys. Lett. {\bf
    44}, 360 (1998).

\bibitem{ankerholdJ} J. Ankerhold, Europhys. Lett. {\bf 67}, 280
  (2004).

\bibitem{strat} R. L. Stratonovich, {\it Topics in the Theory of
    Random Noise} Vol. 1 (Gordon and Breach, New York, 1963).

\bibitem{tilted} R.~L. Stratonovich, Radiotekhnika i elektronika {\bf
    3} (No. 4), 497 (1958); V.~I. Tikhonov, Avtomatika i telmekhanika
  {\bf 20} (No. 9), 1188 (1959); R.~L. Stratonovich, Topics in the
  Theory of Random Noise, Vol. II (Gordon and Breach, New York-London,
  1967); Yu.~M.  Ivanchenko and L.~A.  Zil'berman, Zh. Eksp. Teor. Fiz
  {\bf 55} 2395 (1968) [Sov. Phys. JETP {\bf 28}, 1272 (1969)]; V.
  Ambegaokar and B.~I. Halperin, Phys. Rev. Lett. {\bf 22}, 1364
  (1969); I. Zapata, J. Luczka, F. Sols, P. H\"anggi, Phys. Rev. Lett.
  {\bf 80}, 829 (1998).

\bibitem{rudnicki} J. \L uczka, R. Rudnicki, P. H\"anggi, Physica A
  {\bf 351}, 60 (2005).

\bibitem{PR} P. H\"anggi and H. Thomas, Phys. Rep. {\bf 88}, 207
  (1982); see section $2.4$.

\bibitem{Mironov} W. I. Tikhonov and M. A. Mironov, Markovian
  Processes (Sov. Radio, Moscow, 1977).

\bibitem{peclet} E. P\'eclet, Ann. Chim. Phys.  {\bf 3}, 107 (1841);
  E. P\'eclet, {\it Trait\'e de la Chaleur Consider\'ee dans ses
    Applications}, 3 vols., (Hachette, Paris, 1843).

\bibitem{MachuraJCP} L. Machura, M. Kostur, F. Marchesoni, P. Talkner,
  P. H\"anggi and J.  Luczka, J. Phys.: Condens. Matter {\bf 17},
  S3741 (2005); L. Machura, M. Kostur, P. Talkner, J. Luczka, F.
  Marchesoni, P. Hanggi, Phys. Rev. E {\bf 70}, 061105 (2004).

\end{thebibliography}
\end{document}